\documentclass[12pt]{iopart}

\usepackage{graphicx}

\begin{document}

\title[Nesting and Lifetime Effects]{Nesting and lifetime effects in the FFLO state of quasi-one-dimensional imbalanced Fermi gases}

\author{Heron Caldas}
\address{Departamento de Ci\^encias Naturais, Universidade Federal de S\~ ao Jo\~ao del Rei, \\
36301-160, S\~ao Jo\~ao del Rei, MG, Brazil}
\ead{hcaldas@ufsj.edu.br}

\author{Mucio A. Continentino}
\address{Centro Brasileiro de Pesquisas F\'{\i}sicas \\
Rua Dr. Xavier Sigaud, 150, Urca \\ 
22290-180, Rio de Janeiro, RJ, Brazil}
\ead{mucio@cbpf.br}

%\date{\today}

\begin{abstract}

Motivated by the recent experimental realization of a candidate to the Fulde-Ferrell (FF) and the Larkin-Ovchinnikov (LO) states in one dimensional  (1D)  atomic Fermi gases, we study the quantum  phase transitions in these enigmatic, finite momentum-paired superfluids. We focus on the FF state and investigate the effects of the induced interaction on the stability of the FFLO phase in homogeneous spin-imbalanced quasi-1D Fermi gases at zero temperature. When this is taken into account we find a direct transition from the fully polarized to the FFLO state  in agreement with exact solutions.
Also, we consider the effect of a finite lifetime of the quasi-particles states in the normal-superfluid instability. In the limit of long lifetimes, the lifetime effect is irrelevant and the transition is directly from the fully polarized to the FFLO state. We show, however,  that for sufficiently short lifetimes there is a quantum critical point (QCP), at a finite value of the mismatch of the Fermi wave-vectors of the different quasi-particles, that we fully characterize. In this case the transition is from the FFLO phase to a normal partially polarized state with increasing mismatch.
\end{abstract}

%\pacs{}

%\keywords{Suggested keywords}%Use showkeys class option if keyword display desired
\maketitle

\section{Introduction}

Recently, there has been increased interest in the theory of one-dimensional (1D) imbalanced Fermi systems, partly because of the relevance of these theories for the understanding of the Fulde and Ferrell~\cite{Fulde64} and Larkin and Ovchinnikov~\cite{Larkin64} (FFLO) phase. The FFLO is an exotic phase proposed approximately forty years ago, where atoms of opposite momenta and spins form Cooper pairs with finite momentum. In spite of intense theoretical and experimental efforts, the FFLO phase remains elusive.

In three-dimensional (3D) systems, in the strongly-interacting limit, experiments show~\cite{Exp4,Exp5,Exp6,Exp7,Exp8} that the gas phase separates with an unpolarized superfluid core surrounded by a polarized shell~\cite{Caldas1,Caldas2}, with no evidence for the FFLO phase~\cite{Exp8}. However, in one-dimensional (1D) imbalanced Fermi gases, the observed density profiles~\cite{Hulet} agree quantitatively well with theories that exhibit the 1D equivalent of FFLO correlations at low temperatures~\cite{Yang,Orso,Drummond,Drummond2}. These experimental measurements~\cite{Hulet} of density profiles of a two spin mixture of ultracold $^6{Li}$ atoms trapped in an array of 1D tubes show that at finite spin imbalance, the system phase separates with an inverted phase profile as compared to the 3D case. In these 1D experiments a partially polarized core was observed surrounded by wings composed of either a completely paired or a fully polarized Fermi gas, depending on the degree of polarization. 

This recent experimental observation of what can be seen as a strong candidate for FFLO-like correlations in 1D has motivated  theoretical investigations of possible mechanisms responsible for its stability. The increased stability of FFLO-like phases in 1D can be understood as a nesting effect, where a single wavevector connects all points on the Fermi surface, allowing all atoms on the Fermi surface to participate in finite momentum pairing, while in 3D, only a small portion of these atoms are able to contribute. This 1D Fermi surface nesting enhancement of the instability of the normal to the FFLO state is analogous to the conventional charge density wave (CDW) instability~\cite{Huse}.

In the paring mechanism, besides the particle-particle channel considered by Nozi\' eres and Schmitt-Rink (NSR)~\cite{NSR}, there is a correction of the two-body pairing interactions considered first by Gorkov and Melik-Barkhudarov (GMB)~\cite{GMB61}. This correction accounts for the induced interactions which arise between atoms at the Fermi level due to the polarization of the medium. It has been shown that these induced interactions suppress the superfluid transition temperature by a factor of about $2.22$ in 3D~\cite{Pethick00} and $2.72$ in 2D~\cite{Baranov,Mateus} spin-balanced Fermi gases, respectively, when compared with the mean-field (MF) results. The GMB correction was considered recently in various situations as, for instance, in a spin-balanced Fermi gas in an optical lattice \cite{Kim09,Heiselberg09}, in a homogeneous three-components Fermi gas \cite{Pethick09}, and in the unitary limit of spin-balanced \cite{Yu09} and imbalanced 3D Fermi gases \cite{Yu10}.

In this paper we study the zero temperature (T) phase diagram of a quasi-1D imbalanced Fermi system as a function of the mismatch $h$ between their Fermi wavevectors~\cite{shimahara}.  This is relevant for the nearly 1D Fermi gases we are interested in. We show that including induced interactions through a Random Phase Approximation (RPA) is essential to correct the MF (naive) pairing fluctuations results, since they give rise to a finite critical field (or mismatch) separating a fully polarized phase from the FFLO state, and in this way, to reveal the presence of the nesting effect in spin-imbalanced quasi-1D (ideal) Fermi gases. 

It is possible to conceive in actual physical systems mechanisms by which the quasiparticle states in the normal phase acquire a finite lifetime. For example, due to different types of unknown (or unrecognized) scattering mechanisms not included in the pairing interaction,  or due to an inhomogeneous distribution of the atoms in a trap. In condensed matter systems, for quasi-one dimensional  superconductors, disorder may have a more mundane origin, as defects or impurities~\cite{wosnitza}.  We assume the existence of weak inter-tube interactions so that the effects of localization are not so severe. We show here how this lifetime effect modifies the $T=0$ phase diagram of the 1D gas. In the limit of short lifetimes of the quasiparticle states in 1D we find a quantum phase transition from the normal-to-inhomogeneous  superfluid as the Fermi wave-vector mismatch is reduced from the normal phase. This $T=0$ transition is continuous or second order. It is associated with a QCP at a critical value of the field (mismatch) $h_c$. We fully characterize this QCP obtaining its dynamic quantum critical exponent and universality class.  On the other hand, for sufficiently long lifetimes (weak disorder), lifetime effects turn out to be irrelevant and we recover the previous results of including only induced interactions.
Our results imply that  for sufficiently strong disorder, the region in the phase diagram where the FFLO phase exits is reduced without necessarily being destroyed, even in 1D.  The instability of the normal state that we consider is that for a FFLO superfluid state characterized by a single wave-vector $\mathbf{q}$. This is the first zero temperature instability that occurs as the effective Zeeman field $h$ is reduced \cite{shimahara} from the normal phase. 

The majority of the recent literature on 1D spin-imbalanced Fermi gases considers the simplest possible system that exhibits FFLO-type pairing namely, the Yang-Gaudin model or its lattice version, the Hubbard model with attractive interactions \cite{integra,eta,soliton,vivi,liu}. Both models are exactly solvable (or integrable) and their energy spectra and thermodynamical properties can be calculated exactly using the Bethe ansatz and numerical methods \cite{integra}.
The Yang-Gaudin model does not include inter-tube couplings and eventually it will be necessary to consider these \cite{soliton,lecturenotes}   to fully describe the experiments.

The theoretical predictions and in particular the phase diagram obtained using the integrable Yang-Gaudin model \cite{Orso,Drummond,guan,guanqpt} agree very well with the experimentally observed density profiles and support the stability of the FFLO phase in 1D.  Here we give a robust and transparent physical explanation of this stability as a consequence of nesting effects. We also show explicitly that the medium  indubitably modifies the fermion-fermion interaction g, due to many-body effects an effect which can not be clearly seen in the exact  approaches. We point out that our results including induced interactions are consistent with the exact results since, for long lifetimes, we find a direct transition from the fully polarized to the FFLO state. This is not surprising since the physics of this problem is determined by the nesting effect which is present in both approaches, either explicitly or implicitly.

Our study is based on  calculations of pair and density fluctuations, i.e., of the particle-particle (pair) and particle-hole susceptibilities, respectively, that certainly are present in a Fermi gas, in any dimension. In 1D these two physical quantities are of extreme importance, since both diverge, which are indications of some order in the system. 
As will be shown below, our main results do not depend on the particular Hamiltonian used to describe the attractive fermionic gas. Rather they arise from fundamental quantities, namely the 1D fermionic dispersion relations that always can be linearized close to the Fermi points, regardless of its precise nature, and the subsequent calculation of the particle-hole and particle-particle susceptibilities  \cite{Giamarchi}. Thus, given that it is of fundamental importance to consider the Fermi surface properties at 1D or, the particle-particle and particle-hole interactions near it, our results can be considered as model independent. In this sense, our approach is complementary to those previous studies based on the real space Yang-Gaudin and Hubbard models.

\section{Model Hamiltonian}

To begin, let us consider a non-relativistic dilute (i.e., the particles interact through a short-range attractive interaction) 1D spin-polarized Fermi gas, described by the following single-channel model Hamiltonian

\begin{eqnarray}
\label{eq-1}
{\cal H}&=&H-\sum_{k,\alpha}\mu_{\alpha}n_{\alpha}\\
\nonumber
&=&\sum_{k} {\epsilon}^{a}_k a^{\dagger}_{k} a_{k}+{\epsilon}^{b}_k b^{\dagger}_{k} b_{k}
+ g_{1D} \sum_{k,k'} a^{\dagger}_{k'} b^{\dagger}_{-k'} 
b_{-k} a_{k},\,
\end{eqnarray}
where $a^{\dagger}_{k}$, $a_k$ are the creation and annihilation operators for the $a$ particles (and the same for the $b$ particles) and ${\epsilon}^{\alpha}_k$ are their dispersion relation, defined by ${\epsilon}^{\alpha}_k=\xi_k - \mu_{\alpha}$, with $\xi_k=\hbar^2 k^2/2m$ and $\mu_{\alpha}$  the chemical potential of the non-interacting $\alpha$-species, $\alpha=a,b$. A special case described by this Hamiltonian is that of identical spin $S=1/2$ particles under an external magnetic field $h$. In this case, the $a$ and $b$ correspond to the spin up and spin down bands with their degeneracy raised by the magnetic field. The dispersion relations are then given by $\epsilon_k^{a,b} = \xi_k - \mu_{a,b} = \xi_k - \mu \mp h$. In particular, we will approximate the dispersions around their Fermi energies by $\epsilon_k^{a,b}= v_F (k-k_F) \mp h$, since the relevant states are near the Fermi momenta. $v_F$ is the Fermi velocity. For future notation, we rewrite this equation as $\epsilon_k^{a,b}= v_F (k-k_F^{a,b})$,  where $k_F^{a,b}= k_F \pm h / v_F$ are the Fermi wave-vectors of the different particles or bands.  An important parameter in the present study is the \textit{Fermi wave-vector mismatch}, $\delta k_F= k_F^a-k_F^b= 2h/v_F$. To reflect an attractive (s-wave) interaction between particles $a$ and $b$ we take  $g_{1D} \equiv g <0$. 

\section{The Ideal Case: Infinite Lifetimes}

In this section we consider quasi-1D imbalanced Fermi system as a function of the mismatch $h$ between their chemical potentials in the ideal case, i.e., in the absence of lifetime effects. Our aim is to investigate the quantum phase transition from the normal-to-inhomogeneous FFLO superfluid phase as the Fermi wave-vector mismatch is reduced from the normal polarized state.

\subsection{Ginzburg-Landau Theory and the FFLO Phase}

Since in a homogeneous 1D system the quantum phase transition to the FFLO phase is continuous~\cite{Drummond,Drummond2,Huse}, we expand the action in fluctuations, $|\Delta_{\vec{q}}|$, and obtain \cite{russos,aline},

\begin{equation}
 S_{{eff}}= \sum_{\vec{q}}  \int d \omega_0 \left( \alpha({|\vec{q}|}, \omega_0) |\Delta_{\vec{q}}(\omega_0)|^{2}+
\mathcal{O}\left(|\Delta|^4\right)\, \right),
\end{equation}
where $\alpha(|{\vec{q}}|,\omega_0)=-\frac{1}{g}-\chi_{pp}({|\vec{q}|}, \omega_0)$, with $(\vec{q}, \omega_0)$  being the external momemtum of the particle-particle bubble diagram and $\chi_{pp}(|{\vec{q}}|, \omega_0)$  the pair susceptibility,

\begin{eqnarray}
\label{chi1-1}
\chi_{pp}(|{\vec{q}}|, \omega_0)=
\sum_{\vec{k}} \frac{1-n(\xi^b_{\vec{k}-\vec{q}/2})-n(\xi^a_{\vec{k}+\vec{q}/2})}{\xi^b_{\vec{k}-\vec{q}/2}+\xi^a_{\vec{k}+\vec{q}/2}-\omega_0},
\end{eqnarray}
or
\begin{eqnarray}
\label{chi7-1Dnv3}
\fl \chi(|{\vec{q}}|,\omega_0) =
\frac{m}{4 \pi k_F} \int_{0}^{\omega_c} d \omega \tanh \left(\frac{\omega}{2T}\right) \left[ \frac{1}{\omega + h + (a-\omega_0/2)}+\frac{1}{\omega + h - (a-\omega_0/2)} \right] \\
\nonumber
+ \frac{m}{4 \pi k_F} \int_{0}^{\omega_c} d \omega \tanh\left(\frac{\omega}{2T}\right) \left[ \frac{1}{\omega + h + (a+\omega_0/2)}+\frac{1}{\omega + h - (a+\omega_0/2)} \right],
\end{eqnarray}
where $\omega_c$ is an energy cut-off and $a \equiv v_F |\vec{q}|/2$. In the zero temperature limit,  $\chi_{pp}(|{\vec{q}}|, \omega_0)$ is given by

\begin{eqnarray}
\label{chi7-1Dnv5}
\fl \rm{Re} ~\chi(\bar q,\omega_0) =
N(0) \left[ \ln\left(\frac{\omega_c}{h}\right) -\frac{1}{4} \ln|1-(\bar q + \bar \omega_0)^2|-\frac{1}{4} \ln|1-(\bar q - \bar \omega_0)^2| \right],
\end{eqnarray}
where $N(0)=\frac{m}{\pi k_F}$ is the density of states for both spins at the Fermi energy $E_F=k_F^2/2m$, $\bar q \equiv v_F |\vec{q}|/2h$, $\bar \omega_0 \equiv \omega_0/2h$. Here $v_F$ is the Fermi velocity, and

\begin{eqnarray}
\label{chi7-1Dnv6}
\rm{Im}~\chi(\bar q,\omega_0) =
&-&N(0) \frac{\pi}{2},~{\rm if}~h < (a-\omega_0/2)^2,\\
\nonumber
&-&N(0) \frac{\pi}{4},~{\rm if}~(a-\omega_0/2)^2 < h < (a+\omega_0/2)^2,\\
\nonumber
&0&, ~{\rm if}~h > (a+\omega_0/2)^2.
\end{eqnarray}
The static pair susceptibility reads

\begin{eqnarray}
\label{chi10-1D}
\chi_{pp}(\bar q) &=&
 N(0) \left[ \ln \left(\frac{2 \omega_c}{2 h} \right) - \frac{1}{2} \ln \left(| 1- {\bar q}^2 \right|) \right].
\end{eqnarray}
At the continuous phase transition, we find $\alpha(\bar q)=-\frac{1}{g}-\chi_{pp}(\bar q)=0$, which yields the critical field $h_c(\bar q)$:

\begin{eqnarray}
\label{cf}
\frac{h_c}{\Delta_0}=\frac{1}{2\sqrt{|1-\bar q^2|}},
\end{eqnarray}
where $\Delta_0=2 \omega_c {\rm exp}(-1/N(0)|g|)$ is the zero temperature BCS gap. This expression diverges for $\bar q= \bar q_c=1$ i.e., for $q_c=\frac{2h}{v_F}$, yielding $h_c=\infty$. Since the two species Fermi momenta can be written as $k_F^{b,a}=k_F \pm \frac{h}{v_F}$, their difference is $k_F^b-k_F^a=\frac{2h}{v_F}$, and we find that at the critical mismatch the pair wave vector reads $q_c=k_F^b-k_F^a$. Thus, the  calculation above shows that FFLO type of correlations are so strong in quasi-1D systems with attractive interactions, that differently from d = 2~\cite{PRB} and d = 3~\cite{fflo2d}, in 1D the FFLO phase persists even in the presence of an arbitrarily large magnetic field~\cite{Machida}.  However, we show below that including induced interactions substantially modifies this scenario and there is a finite $h_f$ beyond which FFLO correlations disappear. This field coincides with that at which the system becomes fully polarized in agreement with the exact results \cite{lecturenotes}.

\subsection{Induced Interaction in a Spin Polarized Fermi Gas}

NSR have shown that as the superconducting transition temperature $T_c$ is approached from above, Cooper pair fluctuations grow in amplitude, and the pair susceptibility (which measures the tendency of pairs to form in response to an external pair field) diverges. The pair fluctuation is expressed by the pair susceptibility of Eq.~(\ref{chi1-1}), and NSR showed that $\alpha(|{\vec{q}}|=0,\omega_0=0)=0$ is simply the Thouless condition for weak coupling superconductivity~\cite{NSR}.

Besides the pairing fluctuations that must be taken into account, as pointed out by NSR, to obtain the correct superfluid transition temperature $T_c$ of the BEC-BCS crossover,  there is another effect of particle-hole fluctuations that affects the superfluid state. Namely, there is a change in the coefficient $\alpha(|\vec{q}|)$ due to screening of the interspecies (or induced) interaction, known as the GMB correction~\cite{GMB61}. In the BEC side, the NSR fluctuation is dominant, while the GMB fluctuation becomes weaker towards the BEC side and vanishes in this region due to the disappearance of the Fermi surface~\cite{Mateus}.

In the original work by GMB~\cite{GMB61}, the induced interaction was obtained in the BCS limit by second-order perturbation~\cite{GMB61,Pethick00}. The diagram in Fig.~\ref{rpa}, describes a scattering process in which the conservation of total momentum implies that $p_1+p_2\rightarrow p_3+p_4$. Besides, the frequencies and momentum are set to zero, so that $\vec{p}_1=- \vec{p}_2$ and $\vec{p}_3=-\vec{p}_4$. This leads to the induced interaction $g_{\mathrm{ind}}( p_1, p_4)= - g^2 \chi_{ph}(p_1-p_4)$~\cite{Pethick00}, where $\chi_{ph}(p_1-p_4)$ is the particle-hole susceptibility, $p_{i}=({\bf p}_{i}, \omega_{l_i})$ is a vector in the space of wave-vector ${\bf p}$ and fermion Matsubara frequency $\omega_{l}=(2l+1)\pi/(\hbar\beta)$, $\beta=1/(k_BT)$. The term $g_{\mathrm{ind}}( p_1, p_4)$ describes the modification of the inter-particle interaction $g$ due to many-body effects (i.e., the presence of other particles). 

In general, if the number of electrons in the band is such that the particle-hole susceptibility $\chi_{ph}$ is well-behaved, or in dimensions higher than 1 where $\chi_{ph}$ does not diverge, the second order term in the induced interaction is sufficient to describe the system and calculate the change of the critical parameters. However, for general cases the GMB treatment can be extended to a region where all orders are important, and one has to sum an infinite series of the diagram represented in Fig.~\ref{rpa} in a random phase approximation (RPA). This yields~\cite{Yu09,Yu10}:

\begin{equation}
g_{\mathrm{ind}}( p_1, p_4)= -\frac{g^2 \chi_{ph}(p_1-p_4)}{1+g\chi_{ph}(p_1-p_4)}.
\end{equation}
Including the total induced interaction, the effective interaction between atoms with different spins is given by:

\begin{figure}
\begin{center}
\includegraphics[width=6cm]{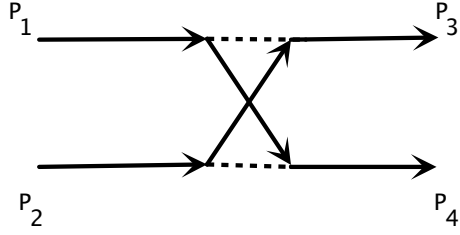}
\caption{ The lowest order particle-hole diagram which generates the induced interaction $g_{\mathrm{ind}}( p_1, p_4)$. Arrowed and dashed lines describe fermionic propagators and the coupling $g$ between the atoms, respectively.}
  \label{rpa}
  \end{center}
\end{figure}

\begin{equation}
g_{\mathrm{eff}}( p_1, p_4) \equiv g_{\mathrm{eff}} = g + g_{\mathrm{ind}}( p_1, p_4)=\frac{g}{1+g\chi_{ph}( p_1, p_4)}.
\label{induce}
\end{equation}
Next, we calculate the polarization function $\chi_{ph}(p_1-p_4) \equiv \chi_{ph}(p')$ for a 1D spin polarized fermionic gas. This is given by,

\begin{equation}
\chi_{ph}(p')= {1\over \hbar^2 \beta {\rm L}}\sum_p
\mathcal{G}_{0,b}(p)\mathcal{G}_{0,a}(p+p')
=\int {\rm{d} {\bf k}\over (2\pi)^2} {f_{{\bf
k}b}-f_{{\bf k}+{\bf k'}a}\over i\hbar
\Omega_l+\epsilon^b_{{\bf k}}-\epsilon^a_{{\bf k}+{\bf k'}}},
\label{chi0}
\end{equation}
where $p'=({\vec k}', \Omega_{l})$, $\Omega_{l}=2l\pi/(\hbar\beta)$ is the Matsubara frequency of a boson, ${\rm L}$ is the size of the system, $f(k)$ is the Fermi distribution function $f({{\cal{E}}_k^{a,b}})=1/(e^{\beta {\cal{E}}_k^{a,b}}+1)$, with $\beta=1/T$, where we have set $k_B=\hbar=1$. The Matsubara Green's function of a non-interacting Fermi gas is given by $\mathcal{G}_{0\sigma}(p)=1/(i\omega_l-\epsilon_{\bf k\sigma})$.  We calculate the polarization bubble in the static limit ($\Omega_l=0$) and find

\begin{equation}
\chi_{ph}(k')=- N(0) f(x,h),
\label{chi1}
\end{equation}
where $k'=|{\vec k}'|$ is equal to the magnitude of $\vec{p}_1+\vec{p}_3= \vec{p}_1-\vec{p}_4$, so $k'=\sqrt{(\vec{p}_1+\vec{p}_3).(\vec{p}_1+\vec{p}_3)}=\sqrt{\vec{p}_1^2+\vec{p}_3^2+ 2\vec{p}_1. \vec{p}_3}=\sqrt{\vec{p}_1^2+\vec{p}_3^2+ 2|\vec{p}_1|| \vec{p}_3|\cos \phi}$, where $\phi$ is the angle between $\vec{p}_1$ and $\vec{p}_3$. Since the scattering is in 1D, the only values of $\phi$ are $0$ or $\pi$. Performing the calculation we obtain the real function $f(x,h)$ in the form of a generalized Lindhard function given by,

\begin{equation}
f(x,h)=\frac{1}{4x} \left[ \ln \left| \frac{1+x^{b}H^{b}}{1-x^{b}H^{b}} \right| + \ln \left| \frac{1+x^{a}H^{a}}{1-x^{a}H^{a}} \right| \right],
\label{chi2}
\end{equation}
where $x=\frac{k'}{2 k_F}$, $x^{b, a}=\frac{k'}{2 k_F^{b, a}}$, $k_F^{b, a}=\sqrt{2m \mu^{b, a}}$, $H^{b, a}=1\pm \frac{4mh}{k'^2}$. Note that at $h=0$, the well known result for $1D$ balanced systems is recovered, 

\begin{equation}
f(x)=\frac{1}{2x} \ln \left| \frac{1+x}{1-x} \right|,
\label{chi3}
\end{equation}
for which $\chi_{ph}(k'=0)= - N(0)$. The function $f(x,h)$ diverges for both $x^{b,a}H^{b,a}=1$ or for either, $x^{a}H^{a}=1$ or $x^{b}H^{b}=1$. Solving, for example,  $x^{b}H^{b}=1$ for $k'$ we find:

\begin{equation}
{k'}^{b}=k_F^{a} + k_F^{b}=2 k_F.
\label{q1}
\end{equation}
Thus, at ${k'}^{b} \equiv k'= k_F^{b}+k_F^{a}=2 k_F$ the function $f(x,h)$ in Eq.~(\ref{chi2})  diverges (a similarly condition holds for ${k'}^{a}$). This corresponds to $\phi=0$ and $|\vec{p}_1|=|\vec{p}_3|=k_F=\sqrt{2M\mu}$, meaning that both scattering particles are at the Fermi surface. Notice that in metallic systems, the divergence of $\chi_{ph}(k')$ for a given value of $g$ is often related to an instability to a charge ordered phase~\cite{Kim09}. The result above shows that even in the presence of an external magnetic field, the particle-hole susceptibility diverges at the same value of $k$, as in the absence of the ``field'' $h$, namely for $k^{\prime}=2 k_F$.

\begin{figure}
\begin{center}
\includegraphics[angle=0,width=7cm]{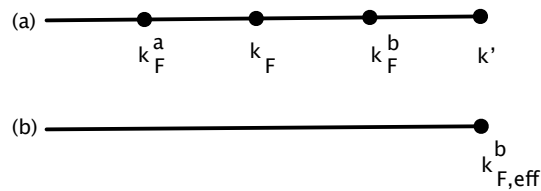}
\caption{ Geometrical illustration of the Fermi surfaces of a 1D imbalanced gas for a given mismatch $\delta k_F = k_F^b- k_F^a$. In (a) we see the Fermi points and wavevectors of the $a$ and $b$ species, relative to $k_F$, and the nesting vector $k^{\prime}=k_F^{a}+k_F^{b}=2k_F$. In (b), the system is fully polarized and non-interacting, due to the induced interaction, with an effective $b$-Fermi surface $k_{F, eff}^b=2 k_F$, with $k_{F}^a=0$. This fully polarized state occurs for a field  $h_f= \mu = \frac{1}{2}v_F k_F$.}
  \label{FP}
  \end{center}
\end{figure}

Considering particle-hole fluctuations, we replace $g$ by $g_{\mathrm{eff}}$ and the instability condition for the superconducting phase is given by:
\begin{equation}
\alpha_{eff}(\bar q)=-1-g_{eff}(k^{\prime})\chi_{pp}(\bar q)=0
\label{g-Gorkov1}
\end{equation}
%where we took $k^{\prime}=q$.
It can be easily verified from this equation that the wave vector for which this condition is first satisfied is still given by $\bar{q}=\bar{q}_c=1$, which gives a critical field $h_c = \infty$. An interesting and new possibility occurs when $k^{\prime}=k_F^{b}+k_F^{a}= 2 k_F = q_c = k_F^{b}-k_F^{a}$. However, this is possible if, and only if, $k_F^{a}=0$. Then, we conclude that the many-body effects brought about by the nesting wave vector $k^{\prime}$ which connects the two Fermi points $k_F^{a}$ and $k_F^{b}$ gives rise a new effective $b$ species Fermi surface, $k_{F, eff}^b=2 k_F$, with $k_F^a=0$. Indeed, it can be seen from the divergence of the particle-hole susceptibility and Eq.~(\ref{induce}) that the effective interaction is $g_{eff}=0$. This situation $q_c=2k_F$, as we just verified, corresponds to the fully polarized gas. This fully polarized gas, which is equivalent to empty the band of $down$ spins and accommodate all in band $b$, such that, $k_F^a=0$ and $k_F^b=2 k_F$, is reached for a field $h_f= \mu = \frac{1}{2}v_F k_F$, as illustrated in Fig.~\ref{FP}. Then for $h \ge h_f$ the fully polarized system can be considered as non-interacting and it remains normal for $h \ge h_f$. For $h<h_f$ the system enters the FFLO phase.

The spin polarization is defined as $P=\frac{n^b-n^a}{n^b+n^a}$, where $n^{a,b}$ are the number densities. Since in a 1D system we have $k_F^{a,b}=\frac{\pi}{2}n^{a,b}$, the polarization can be written as:

\begin{equation}
P=\frac{(\mu+h)^{1/2}-(\mu-h)^{1/2}}{(\mu+h)^{1/2}+(\mu-h)^{1/2}}.
\label{P}
\end{equation}
For $h_f=\mu$, we have $P=1$ as expected.

As we mentioned at the Introduction Section, exact results are obtained for 1D imbalanced Fermi gases using the Bethe Ansatz within the Gaudin-Yang model. See, for instance~\cite{Orso,lecturenotes}, where the ground state energy expression for a homogeneous system is given in terms of spectral functions, which in turn are solutions of two coupled integral equations. The main results presented in~\cite{Orso,lecturenotes} for a homogeneous 1D imbalanced Fermi gas with fixed total density $n=n_{\downarrow}+n_{\uparrow}$ and density difference $s=n_{\uparrow}-n_{\downarrow}$, with $0\leq s \leq n$ are: for $s=0$, the ground state of the system is a fully paired (BCS) state. For $s=n$ the system is a fully polarized gas consisting of solely $\uparrow$ fermions. And finally, for any $0<s<n$ the gas is partially polarized and is a superfluid of the FFLO type. These are exactly the same results we have obtained, as a manifestation of the nesting effect. We stress that since the nesting effect is a intrinsic and universal phenomena in 1D Fermi systems~\cite{Giamarchi,lecturenotes} it should be properly considered, as we did here.

We conclude this section with the result that in the ideal case, where the quasi-particles have an infinite lifetime, the FFLO phase will occur for all $h<h_f$ and for any strength of the attractive interaction. Here $h_f$ is the field above which the system is fully polarized. The quantum phase transition in this case is directly from the fully polarized state to a phase with FFLO correlations \cite{lecturenotes}. The nature of the quantum phase transitions in the pure case of infinite lifetime have been investigated by Guan and Ho~\cite{guanqpt} and at least for the case they are driven by changes in the chemical potential they belong to the universality class of density-driven transitions with dynamic exponent $z=2$ and $\nu=1/2$ as in the case of the repulsive 1d Hubbard model~\cite{livro,modugno}.

\section{Lifetime effects}

In this section we study lifetime effects (LT) in the phase diagram of 1D attractive imbalanced Fermi gases. In cold atom systems the trap to confine the atoms gives rise to an inhomogeneous atomic distribution, which can be described, for example,  by a chemical potential, which depends on the distance from the center of the trap. In this case, since translation invariance is broken, the momentum or wave-vector $k$ is not the good quantum number to describe the quasi-particle states in the trap. However, it is still convenient to use this representation, in which case it is appropriate to introduce a finite lifetime to these states. In the quasi-one-dimensional organic superconductors~\cite{wosnitza}, life-time effects arise from impurities or defects.

Taking into account the finite lifetime of the quasi-particle states in the momentum representation, the particle-particle dynamic susceptibility can be written as, \cite{desordem} 
\small{
\begin{eqnarray}
\fl \chi(q,\omega_0)=\frac{N(0)}{4} \int_{0}^{\omega_c} dx  \Bigg[ \left (\frac{1}{x+h+v_F q/2 -\omega_0/2 + i \gamma/2}+\frac{1}{x+h-v_F q/2 + \omega_0/2 - i \gamma/2}\right)   \nonumber \\
+ \left( v_F q/2 \rightarrow -v_F q/2 \right) \Bigg],
\end{eqnarray}}
where $\gamma=\tau^{-1}$ is the inverse of the lifetime of a quasiparticle $q$-state in the normal phase.
This approach is formally similar to that used to investigate the effect of non-magnetic impurities in higher dimensional (3D and 2D) FFLO superconductors
\cite{takada,mineev}. In this case the main interest was to obtain the reduction in the critical temperature of the superconductor. Here we will concentrate in the zero temperature phase diagram of the 1D system and will be able to fully characterize the new QCP that arises due to the lifetime effects.

\begin{figure}
\begin{center}
\includegraphics[width=7cm]{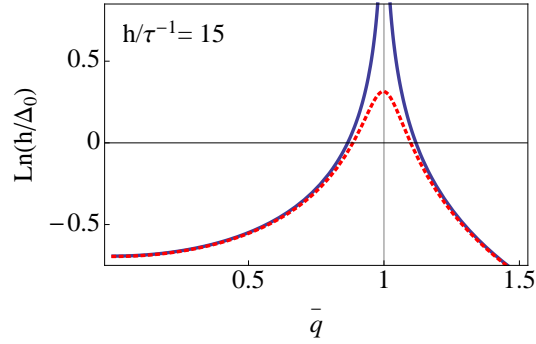}
\end{center}
\caption{(Color online) Critical magnetic field for the normal-to-helicoidal superconductor quantum phase transition as a function of normalized wave-vector for pure (full line) and homogeneous, with finite lifetime, (dashed line) quasi-one dimensional systems.}
\label{q22}
\end{figure}

Since we are interested in studying the effect of this finite lifetime on quantum criticality, we start calculating the real part of the static particle-particle susceptibility
\small{
\begin{equation}
\fl \Re e \chi(q,\omega_0=0)=\frac{N(0)}{2} \int_{0}^{\omega_c} dx  \left[\frac{x+h+v_F q/2}{(x+h+v_F q/2)^2+ \gamma^{2}/4}+\frac{x+h-v_F q/2}{(x+h-v_F q/2)^2 +\gamma^{2}/4}\right].
\end{equation}}
This can be easily integrated and for $\omega_c \tau >>1$, we get,
\begin{equation}
\Re e \chi(q,\omega_0=0)=N(0) \ln \frac{\omega_c}{h} - \frac{N(0)}{4} \ln  \left[(1+\bar{q})^2+\frac{\gamma^{2}}{4h^2}\right]\left[(1-\bar{q})^2+\frac{\gamma^{2}}{4h^2}\right].
\end{equation}
The condition for the divergence of the interacting pair susceptibility becomes,
\begin{equation}
g_{r} N(0) \left[ \ln \frac{2h}{\Delta_0} + \frac{1}{4} \ln  \left[(1+\bar{q})^2+\frac{\gamma^{2}}{4h^2}\right]\left[(1-\bar{q})^2+\frac{\gamma^{2}}{4h^2}\right]\right]=0,
\end{equation}
where $g_{r}$ is an effective (renormalized by the GMB correction) coupling constant. This condition is first satisfied when the argument of the logarithmic is maximum, that is, for $\bar{q}_c=\sqrt{|1- \gamma^{2}/4h^2|}$. The critical field is given by $h_c=h(\bar{q}_c)$ where,
\begin{equation}
\label{cfd}
\frac{2h(\bar{q})}{\Delta_0} =\frac{1}{\left\{\left[(1+\bar{q})^2+\frac{\gamma^{2}}{4h^2}\right]\left[(1-\bar{q})^2+\frac{\gamma^{2}}{4h^2}\right]\right\}^{1/4}
}.
\end{equation}
Substituting for $\bar{q}_c$ we finally get, 
\begin{equation}
\label{D_0}
h_c=h(\bar{q}_c)=\Delta_0^2/4 \gamma,
\end{equation}
implying that in quasi-1D imbalanced Fermi systems, for sufficiently short lifetimes of the quasiparticle states, the FFLO phase  appears only below a critical mismatch $h<h_c=\Delta_0^2/4 \gamma$ (see figure \ref{q22}).  

Taking into account the results of the previous section we can summarize our results as:
\begin{itemize}
\item If $h_c > h_f$, where $h_f$ was calculated in the previous section, the lifetime effect is irrelevant and the phase transition occurs at $h_f$ directly from the fully polarized state to the FFLO phase.
\item For  short lifetimes,  or strong disorder, such that,  $h_c < h_f$, the transition to the FFLO phase occurs from a partially polarized state. In this case there is an intervening normal partially polarized (PP) phase between the fully polarized (FP) state and the FFLO phase ($h_c<h<h_f$). 
\end{itemize}
We emphasize that the results above do not arise from a MF calculation, but rather on the properties of the pair and density fluctuations that go beyond the MF approximation. Besides, they are not dependent on the Hamiltonian given by~Eq.(1), and rely on the 1D fermionic dispersion relations that can always be linearized close to the Fermi points~\cite{Giamarchi}.

For completeness, it is interesting to apply the Thouless criterion to determine the boundary of the homogeneous BCS phase in the case of small field or mismatch. For this it is sufficient to take $q=0$ in Equations~(\ref{cf}) and~(\ref{cfd}), which yields $h_{c,MF}^0=\Delta_0/2$~\cite{Orso,lecturenotes}, and $h_{c,LT}^0 \approx (\Delta_0/2)(1/(1-(\gamma/\Delta_0)^4)^{1/4}$, for the pure (MF) and disordered cases, respectively. When the induced interactions are considered, $h_{c,GMB}^0=\tilde \Delta_0/2$, where the value of $\Delta$ is reduced compared to the bare case as $\tilde \Delta_0=2 \omega_c {\rm exp}(-1/N(0)|g_{eff}|)$, and $g_{eff}=g/(1-gN(0))$ is the interaction corrected by the GMB correction. Thus, the ``hierarchy' of the critical fields on the boundary of the BCS phase may be described as $h_{c,GMB}^0\leq h_{c,LT}^0<h_{c,MF}^0=\Delta_0/2$.

\begin{figure}
\begin{center}
\includegraphics[width=12cm]{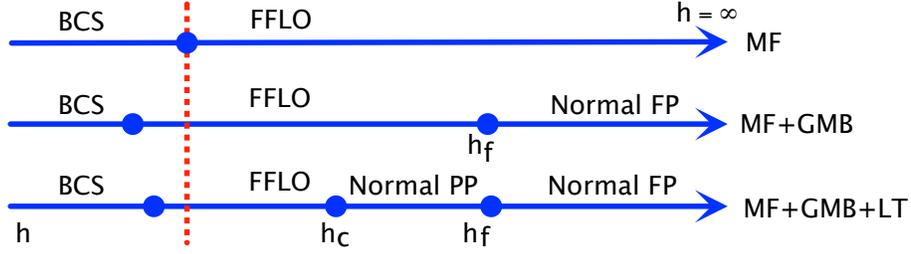}
\end{center}
\caption{(Color online) Zero temperature phase diagram (schematic) as a function of the magnetic field for the different approximations used here (see text).}
\label{fig4}
\end{figure}

A zero temperature phase diagram showing the different phases as the field is increased is shown in Fig.\ref{fig4}, for the different cases studied here.

\section{Nature of the transition at $h_c <  h_f$}

From here on, we will consider the case $h_c < h_f$ and study the quantum phase transition from the normal PP state to the FFLO phase. The real part of the dynamic susceptibility is obtained as,

\begin{eqnarray}
\fl \Re e \chi(q,\omega_0)= N(0) \ln \frac{2 \omega_c}{2h}- \frac{N(0)}{8} \ln\left[(1+\bar{q} -\bar{\omega_0})^2+ \frac{\gamma^{2}}{4h^2}\right]\left[(1-\bar{q} -\bar{\omega_0})^2+ \frac{\gamma^{2}}{4h^2}\right] \\ \nonumber
- \frac{N(0)}{8} \ln\left[(1-\bar{q} +\bar{\omega_0})^2+ \frac{\gamma^{2}}{4h^2}\right]\left[(1+\bar{q} +\bar{\omega_0})^2+ \frac{\gamma^{2}}{4h^2}\right].
\end{eqnarray}
Expanding close to $\bar{q}=\bar{q}_c$ and $\omega_0=0$, we get
\begin{equation}
1 - g_{r} \Re e \chi(q,\omega_0) \approx g_{r} N(0) \left[ \ln \frac{h}{h_c} +(\frac{h \bar{q}_c}{ \gamma})^2 (\bar{q}-\bar{q}_c)^2 +(\frac{h \bar{q}_c}{ \gamma})^2 \bar{\omega_0}^2 \right].
\end{equation}
The imaginary part of the dynamic susceptibility is given by,
\begin{eqnarray}
\fl \Im m \chi(q,\omega_0)=\frac{N(0)}{4} \int_{0}^{\omega_c} dx  \Bigg[\frac{\gamma/2}{\Bigg[x -(\omega_0/2 -h-v_F q/2) \Bigg]^2+\gamma^2/4}+  \\ \nonumber
 \frac{\gamma/2}{\Bigg[x -(\omega_0/2 -h+v_F q/2)\Bigg]^2+\gamma^2/4}+(\omega_0 \rightarrow -\omega_0) \Bigg],
\end{eqnarray}
which can be easily integrated to give,
\begin{equation}
\fl \Im m \chi(q,\omega_0)=-\frac{N(0)}{4} \left[ \tan ^{-1}\left(\frac{2 (\bar{q}-\bar{\omega_0 })}{\bar{\gamma}  \left(1-\frac{(\bar{q}-\bar{\omega_0} )^2-1}{\bar{\gamma}
   ^2}\right)}\right)-\tan ^{-1}\left(\frac{2 (\bar{q}+\bar{\omega_0 })}{\bar{\gamma}  \left(1-\frac{(\bar{q}+\bar{\omega_0} )^2-1}{\bar{\gamma}
   ^2}\right)}\right)\right],
   \end{equation}
where we recall, $\bar{q}=v_F q/2h$, $\bar{\omega_0}=\omega_0/2h$ and we defined $\bar{\gamma}=\gamma/2h$.

Expanding close to $\omega_0=0$, $\bar{q}=\bar{q}_c$ we obtain:
\begin{equation}
\Im m \chi(q,\omega_0)=N(0) \frac{h}{ \gamma} \bar{\omega_0}.
   \end{equation}
Then in the limit $\omega_0 \rightarrow 0$, the frequency dependent part of the imaginary susceptibility dominates over the real part. 

As concerns its quantum critical behavior, the zero temperature phase transition from the normal to the inhomogeneous superconductor state of the homogeneous quasi-1D system in the presence of a finite lifetime of the quasi-particle states can be described by the following effective action, at the Gaussian level \cite{aline},
\begin{equation}
S_{eff} = \int dQ \int d \omega_0 \left[ \delta + Q^2 + |\omega_0| \right]|\Delta(Q, \omega_0)|^2,
\end{equation}
where $\delta = h-h_c$, and $Q=q-q_c$. The quantum critical point associated with this phase transition has a dynamic exponent $z=2$, such that, its effective dimension, $d_{eff} =d+z=3$ \cite{mac,livro}. Consequently, we expect the superfluid transition in the 1D system in the presence of a finite lifetime to be in the universality class of the 3D XY model \cite{livro} due to the two-component nature of the superfluid order parameter.  Effects of temperature \cite{temp} can also be obtained from knowledge of the critical exponents of the quantum critical point. In this case the finite temperature critical line is obtained as $T_c \propto |h-h_c|^{\nu z}$ where $\nu \approx 2/3$ is the correlation length exponent of the 3d-XY model \cite{3dxy} and $z=2$ as obtained previously. Since there is no long range magnetic order in $d=1$ at finite temperatures, this line in practice provides the temperature scale below which the FFLO correlations become important and this varies with the distance to the QCP. Since $(\partial T_c/\partial h)_{h=h_c} = 0$, the characteristic temperature $T_c$   turns out to be very small near the critical field. 

The identification of the universality class of the lifetime induced QCP as being that of the 3d XY model also allows to obtain the behavior of the correlation function of the FFLO fluctuations at $T=0$. Using that the exponent $\eta$ for the order parameter correlation function of the 3d XY model takes the value $\eta=0.0381$ \cite{hasen}, we find that at the QCP the FFLO correlation function decays with distance $r$, as $G(r) \propto 1/r^{d+z-2+\eta}$ \cite{livro}, i.e., $G(r)=1/r^{1.038}$. This exponent turns out to be small or of the same order of that obtained numerically for these type of correlations using the Bethe-ansatz (see Refs.~\cite{eta} and \cite{lecturenotes}).

\section{Discussion and Conclusion}

In spite of theoretical predictions and  intensive experimental activity, the FFLO phase remains elusive.  Motivated by experimental results in cold atom systems and aiming to understand the reasons for the difficulty in observing this phase, we have carried out a detailed study of a FFLO phase in 1D systems, which provide the most favorable conditions for the appearance of this phenomenon. 
We have considered the effect of the induced interaction in an ideal gas to show how the result which predicts long range FFLO correlations in the presence of arbitrarily large magnetic field or mismatches is modified leading to a finite critical field $h_f$. At this field, a nesting condition is satisfied and long range FFLO correlations set in. In agreement with the exact results, the system at $h_f$ goes directly from a regime with strong FFLO  correlations ($h<h_f$)  to the fully polarized normal phase ($h>h_f$).  

In systems, with additional interactions not included in the pairing Hamiltonian, or in the presence of artificial disorder,   a finite lifetime of the quasi-particle excitations may be considered.  We have shown here how the lifetime  effect modifies the $T=0$  phase diagram in 1D. It gives rise to a new characteristic or critical field $h_c$ which depends on the $h=0$ BCS gap and on the lifetime, $\tau=1/\gamma$ of the states (Eq.~\ref{D_0}). If disorder is weak, such that, $h_c>h_f$, disorder is irrelevant and the transition with increasing field is from the FFLO phase to the fully polarized system at $h_f$. However, for strong disorder ($h_c < h_f$) there is a new QCP in the system that we have fully characterized. In this case with increasing mismatch the system goes from the FFLO phase to a normal partially polarized phase at $h=h_c$ and finally to a fully polarized phase at $h_f$. We have fully characterized the QCP at $h_c$. In this case the region  of the phase diagram where the FFLO phase appears is reduced.  

We have also applied the Thouless criterion to determine the boundary of the homogeneous BCS phase in the case of small field $h_c^0$ in the various approaches we considered namely, MF, MF corrected by induced interactions (GMB), and MF considering lifetime effects (LT). We have found that the transition to the BCS phase occur at $h_{c,GMB}^0\leq h_{c,LT}^0<h_{c,MF}^0=\Delta_0/2$.

We hope our results will stimulate further experiments to confirm unambiguously the existence of the FFLO phase in quasi-1D imbalanced Fermi gases.
\newline

Note added on proof. After the completion of this work, we became aware of recent papers that investigated imbalanced fermionic superfluids in arrays of 1D tubes, allowing inter-tube tunneling~\cite{inter1,inter2}.  They found that the evolution of the physical properties between 1D and 3D (including the inverted phase profiles) can be well described the at a MF level. It would be very interesting to consider both the effects of induced interactions and finite lifetime of the particles in the normal phase in the systems considered in the references above.

%\section*{ACKNOWLEDGMENTS}

\subsection*{Acknowledgments}
H. C. and M. A. C. are partially supported by CNPq. The authors also acknowledge partial support from FAPEMIG and FAPERJ. H. C. acknowledges the kind hospitality at CBPF where part of this work was done.

\end{document}